\documentclass[aps,prl,twocolumn,showpacs,preprintnumbers,amsmath,amssymb,superscriptaddress]{revtex4}%

\usepackage{graphicx}%
\usepackage{dcolumn}
\usepackage{amsmath}
\usepackage{color}
\usepackage{multirow}

\begin{document}


\title{SrPt$_3$P: two-band single-gap superconductor}

\author{R.~Khasanov}
 \email[Corresponding author: ]{rustem.khasanov@psi.ch}
\author{A.~Amato}%
\author{P.~K.~Biswas}%
\author{H.~Luetkens}%
\affiliation{Laboratory for Muon Spin Spectroscopy, Paul Scherrer
Institute, CH-5232 Villigen PSI, Switzerland}
\author{ N.~D. Zhigadlo}
\author{B. Batlogg}
\affiliation{Laboratory for Solid State Physics, ETH Zurich, 8093 Zurich, Switzerland}
%

%
%
%

\begin{abstract}
The magnetic penetration depth ($\lambda$) as a function of applied magnetic field and temperature  in SrPt$_3$P($T_c\simeq8.4$~K) was studied by means of muon-spin rotation ($\mu$SR). The dependence of $\lambda^{-2}$ on temperature suggests the existence of a single $s-$wave energy gap with the zero-temperature value $\Delta=1.58(2)$~meV. At the same time $\lambda$ was found to be strongly field dependent which is the characteristic feature of the nodal gap and/or multi-gap systems. The multi-gap nature of the superconduicting state is further confirmed by observation of an upward curvature of the upper critical field. This apparent contradiction would be resolved with SrPt$_3$P being a two-band superconductor with equal gaps but different coherence lengths within the two Fermi surface sheets.
\end{abstract}
\pacs{74.72.Gh, 74.25.Jb, 76.75.+i}

\maketitle


After the discovery of first Fe-based superconductors enormous efforts were made in order to improve their superconducting properties. The intensive search lead to discovery series of new Fe-based materials (see {\it e.g.} Ref.~\onlinecite{Johnston10} for review and references therein) and related compounds such as BaNi$_2$As$_{2}$ \cite{Bauer08}, SrNi$_2$As$_2$ \cite{Ronnig08}, SrPt$_2$As$_2$ \cite{Kudo10}, SrPtAs \cite{Elgazzar12}, without Fe and relatively low superconducting transition temperatures $T_c$'s.

Recently, Takayama {\it et al.} \cite{Takayama12} reported the synthesis of a new family of ternary platinum phosphide superconductors with the chemical formula APt$_3$P (A = Sr, Ca, and La) and $T_c$'s of 8.4, 6.6 and 1.5~K, respectively.
Theoretical studies on the pairing mechanism in these new compounds achieved partially contradicting results \cite{Chen12,Subedi12}.
The authors of Ref.~\onlinecite{Chen12} performed first-principles calculations and proposed that superconductivity is caused by the proximity to a dynamical charge-density wave instability, and that a strong spin-orbit coupling leads to exotic pairing in at least LaPt$_3$P. In contrast, the first principal calculations and Migdal-Eliashberg analysis performed by Subedi {\it et al.} \cite{Subedi12} suggest conventional phonon mediated superconductivity.
Also experimentally seemingly contradicting results were obtained. Based on the observation of nonlinear temperature behavior of the Hall resistivity, the authors of Ref.~\onlinecite{Takayama12} suggest multi-band superconductivity in these new compounds. Note that the presence of two bands crossing the Fermi level was indeed confirmed by ab-initio band structure calculations presented in \cite{Nekrasov12, Kang12,Chen12, Subedi12}. On the other hand the specific heat data of SrPt$_3$P were found to be well described within a single band, single $s-$wave gap approach with the zero-temperature gap value of $\Delta=1.85$~meV \cite{Takayama12}.

In this paper we report on the results of muon-spin rotation ($\mu$SR) studies of the magnetic penetration depth ($\lambda$) as a function of temperature and magnetic field of the novel superconductor SrPt$_3$P. Below $T\simeq T_c/2$ the superfluid density ($\rho_s\propto\lambda^{-2}$) becomes temperature independent which is consistent with a fully gapped superconducting state. The full temperature dependence of $\rho_s(T)$ is well described within a single $s$-wave gap scenario with the zero-temperature gap value $\Delta=1.58(2)$~meV. On the other hand, $\lambda$ was found to increase with increasing magnetic field as is observed in multi-band superconductors or superconductors with nodes in the energy gap. The upper critical field demonstrates a pronounced upward curvature thus pointing to a multi-band nature of the superconducting state of SrPt$_3$P. Our results indicate that SrPt$_3$P is a two-band superconductor with equal gaps but different coherence length parameters $\xi_i$ within two Fermi surface sheets.

\begin{figure}[htb]
\includegraphics[width=1.07\linewidth]{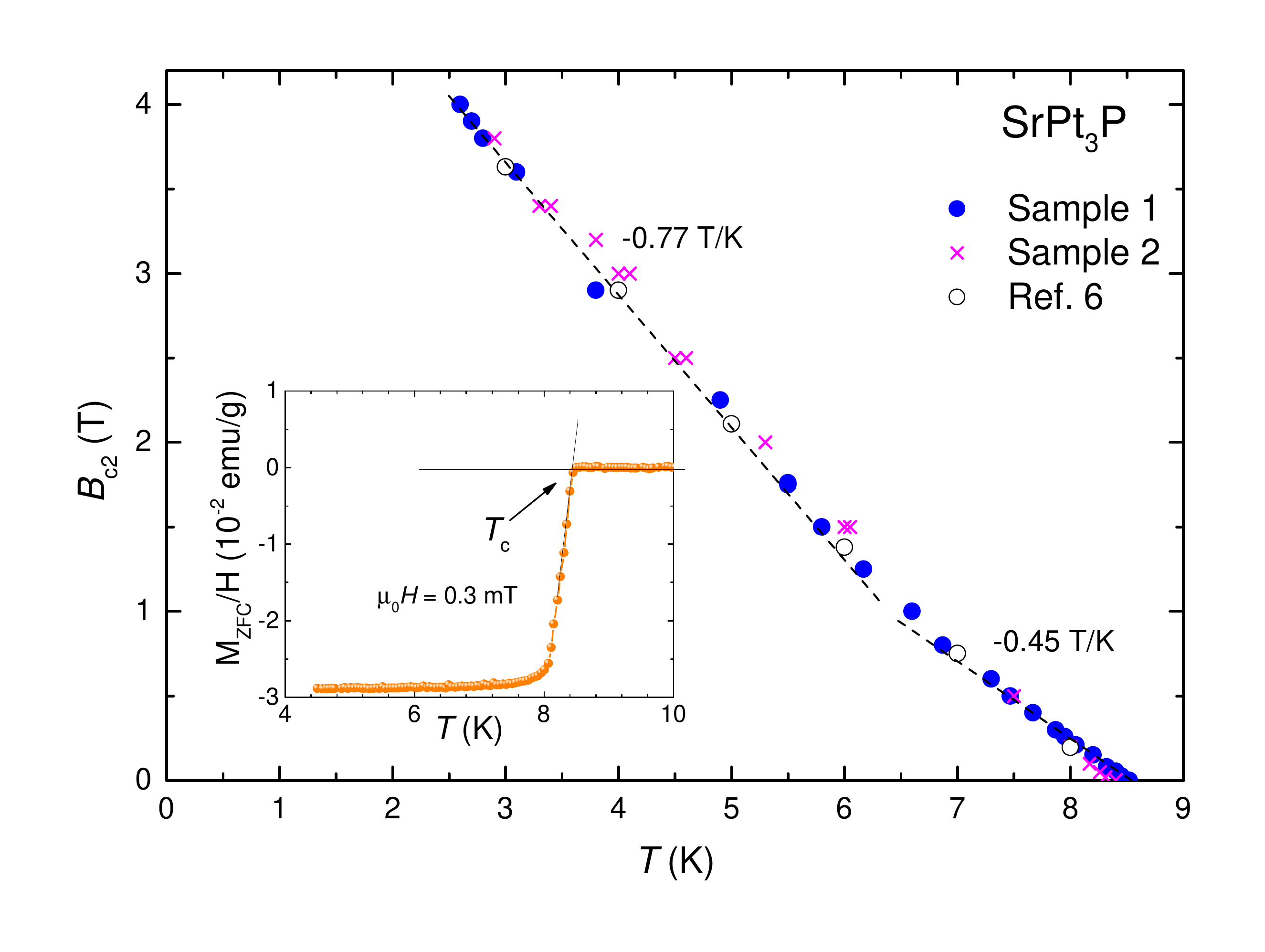}
\caption{(Color online) The temperature dependence of the upper critical field $B_{c2}$ of SrPt$_3$P. The crosses and the circles correspond to two different samples. The solid lines are linear fits of $B_{c2}(T)$ in the vicinity of $T_c$ and for $T\leq$6~K. Open circles are $B_{c2}(T)$ data points from Ref.~\onlinecite{Takayama12}. The inset shows the temperature dependence of the zero field-cooled magnetization $M_{ZFC}$ measured at $\mu_0H=0.3$~mT. }
 \label{fig:Hc2}
\end{figure}

The sample preparation and the magnetization experiments were performed at the ETH-Z\"{u}rich. Polycrystalline samples of SrPt$_3$P were prepared using cubic anvil high-pressure and high-temperature technique. Coarse powders of Sr, Pt, and P elements of high purity ( 99.99\%) were weighed according to the stoichiometric ratio 1:3:1, thoroughly grounded, and enclosed in a boron nitride container, which was placed inside a pyrophyllite cube with a graphite heater. All procedures related to the sample preparation were performed in an argon-filled glove box. In a typical run, a pressure of 2~GPa was applied at room temperature. While keeping the pressure constant, the temperature was ramped up in 2~h to the maximum value of   1050~$^{\rm o}$C, maintained for 20-40~h, and then decreased to room temperature in 1~h. Afterwards, the pressure was released, and the sample was removed. All high-pressure prepared samples demonstrate large diamagnetic response with the superconducting transition temperature of $\simeq$8.4~K (see the inset in Fig.~\ref{fig:Hc2}). The powder x-ray diffraction patterns are consistent with those reported in Ref.~\onlinecite{Takayama12}.

Measurements of the upper critical field $B_{c2}$ were performed  using a Quantum Design 14~T PPMS. The temperature dependence of $B_{c2}$ was obtained from zero field-cooled magnetization curves [$M_{ZFC}(T)$]  measured in constant magnetic fields ranging from 0.3 mT to 4~T (see  Fig.~\ref{fig:Hc2}). For each particular field the corresponding superconducting transition temperature $T_c(B)$ was taken as an intersect of the linearly extrapolated $M_{ZFC}(T)$ curve in the vicinity of $T_c$ with $M_{ZFC}=0$ line (see the inset it Fig.~\ref{fig:Hc2}). 
$B_{c2}(T)$ curve exhibits a pronounced upward curvature  around $\sim6-6.5$~K. Linear fits of $B_{c2}(T)$ in the vicinity of $T_c$ and for $T\leq$6K yield ${\rm d}B_{c2}/{\rm d}T=-0.45$ and -0.77~T/K, respectively. Open circles correspond to $B_{c2}(T)$ data points from Ref.~\onlinecite{Takayama12}. They are in perfect agreement with our data thus implying that the upturn on $B_{C2}(T)$ reported here is  indeed a generic property of SrPt$_3$P compound.
Note that an upward curvature of $B_{c2}(T)$ was also observed previously for a number of materials such as Nb \cite{Williamson_70_Nb-V,Weber_91_Nb}, V \cite{Williamson_70_Nb-V}, NbSe$_2$ \cite{Toyota_76_NbSe,Sanchez_95_NbSe, Zehetmayer_10_NbSe}, MgB$_{2}$ \cite{Zehetmayer_02_MgB2, Sologubenko_02_MgB2, Lyard_02_MgB2}, borocarbides and nitrides \cite{Metlushko_97_BC, Shulga_98_BC, Manalo_01_BC}, heavy fermion systems \cite{Measson_04_HF}, various iron-based \cite{Hunte_08_Fe, Jaroszynski_08_Fe, Ghannadzadeh_14_Fe} and cuprate superconductors \cite{Kortyka_10_Cup, Charikova_13_Cup} and was often associated with two-band superconductivity.

\begin{figure}[htb]
\includegraphics[width=1.03\linewidth]{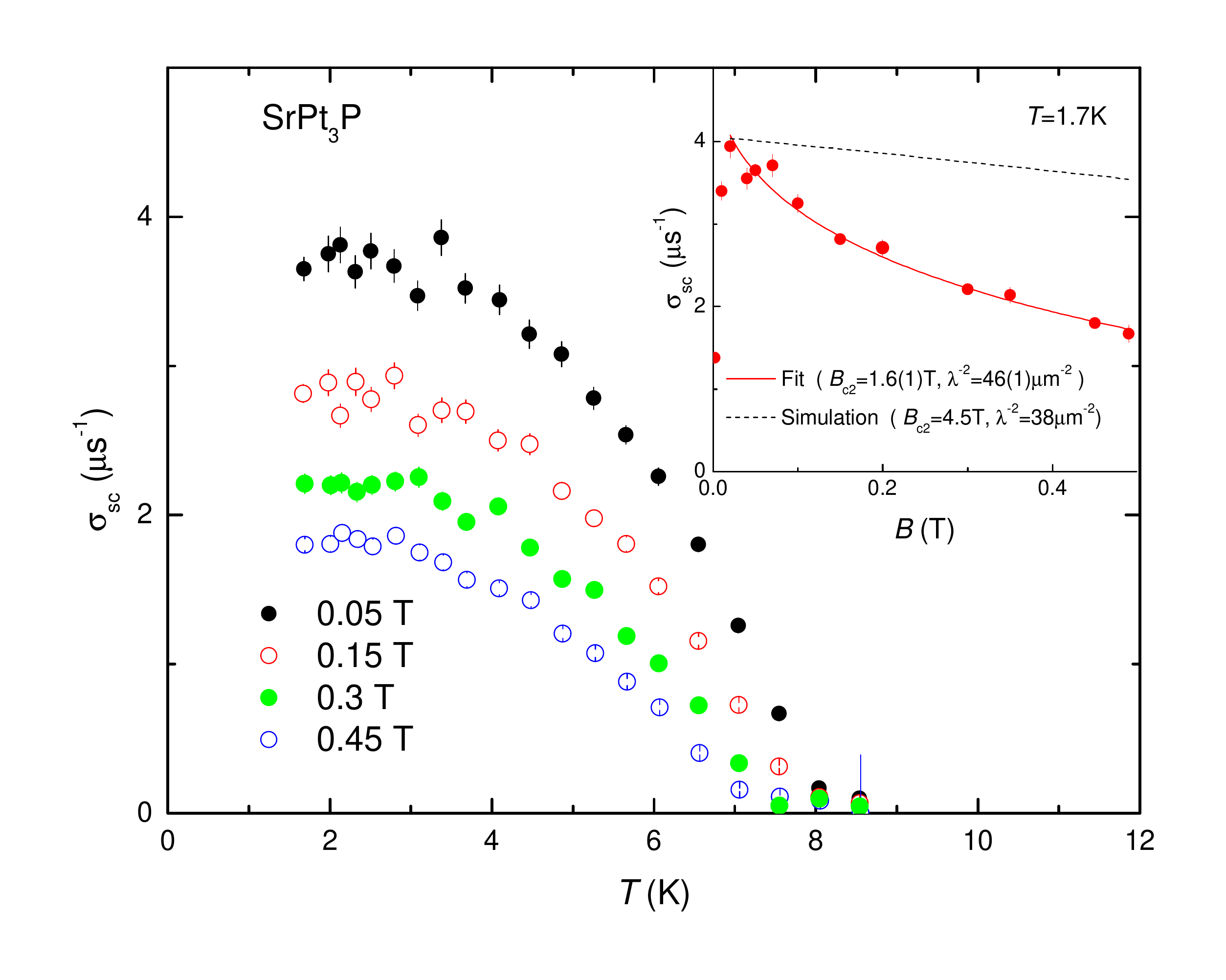}
\caption{(Color online) The temperature dependence of the
depolarization rate $\sigma_{sc}$ caused by formation of FLL in
SrPt$_3$P in fields of 0.05, 0.15, 0.3, and 0.45~T. The inset shows
the dependence of $\sigma_{sc}$ on the applied field $B$ at
$T=1.7$~K. The red solid line is the fit of
Eq.~(\ref{eq:Brandt_sigma_vs_H}) to $\sigma_{sc}(B)$ data with
$\lambda^{-2}=46(1)$~$\mu$m$^{-2}$, $B_{c2}=1.6(1)$~T. The dashed
black line represent $\sigma_{sc}(B)$ as expected for
$\lambda^{-2}=38(1)$~$\mu$m$^{-2}$ and $B_{c2}=4.5$~T obtained in
magnetization experiments.  }
 \label{fig:sigma_vs_T}
\end{figure}

The temperature and the magnetic field dependence of  the magnetic penetration depth $\lambda$ were obtained from transverse-field (TF) $\mu$SR data \cite{Yaouanc11}. The experiments were carried out at the $\pi$E1 beam line at the Paul Scherrer Institute (Villigen, Switzerland). The data were analyzed using the free software package \textsc{MUSRFIT} \cite{Suter12}.
In a polycrystalline sample the magnetic penetration depth $\lambda$ can be extracted from the Gaussian muon-spin depolarization rate $\sigma_{sc}(T)\sim\lambda^{-2}$, which reflects the second moment ($\sigma^2_{sc}/\gamma_\mu^2$, $\gamma_\mu$ is the muon giromagnetic ratio) of the magnetic field distribution due to the flux-line lattice (FLL) in the mixed state \cite{Brandt88, Khasanov06, Khasanov07}. The TF-$\mu$SR data were analyzed using the asymmetry function:
\begin{eqnarray}
A(t)&=& A_{sc} \exp[-(\sigma_{sc}^2+\sigma_n^2)t^2/2]
\cos(\gamma_{\mu}B_{sc} t+\phi)  \cr\cr
 &&+A_b\exp(-\sigma_b^2t^2/2)
\cos(\gamma_{\mu}B_b t+\phi)
 \label{eq:Asymmetry}
\end{eqnarray}
The first term of Eq.~(\ref{eq:Asymmetry}) represents the response of the superconducting part of the sample. Here $A_{sc}$ denotes the initial asymmetry; $\sigma_{sc}$ is the Gaussian relaxation rate due to the FLL; $\sigma_n$ is the contribution to the field distribution arising from the nuclear moment and which is found to be temperature independent, in agreement with the ZF results (not shown); $B_{int}$ is the internal magnetic field sensed by the muons and $\phi$ is the initial phase of the muon-spin ensemble. The second term with the initial asymmetry $A_b$, small $\sigma_b < 0.3$~$\mu$s$^{-1}$ and $B_b$ close to the applied  field corresponds to the background muons stopping in the cryostat and in nonsuperconducting parts of the sample.

Figure~\ref{fig:sigma_vs_T} shows the temperature dependence of $\sigma_{sc}$ in four different fields 0.05, 0.15, 0.3, and 0.45~T. As expected, $\sigma_{sc}$ is zero in the paramagnetic state and starts to increase below the corresponding $T_c(B)$. Upon lowering $T$, $\sigma_{sc}$ increases gradually reflecting the decrease of the penetration depth $\lambda$ or, correspondingly, the increase of the superfluid density $\rho_s\propto\lambda^{-2}$. The overall decrease of $\sigma_{sc}$  with increasing applied field is partially caused by the decreased width of the internal field distribution upon approaching $B_{c2}$. In order to quantify such an effect, one can make use of the numerical Ginzburg-Landau model, developed by Brandt \cite{Brandt03}. This model predicts the magnetic field dependence of the second moment of the magnetic field distribution,  {\it i.e.} $\mu$SR depolarization rate:
\begin{eqnarray}
 \sigma_{sc}[\mu{\rm s}^{-1}]&=&4.83\cdot10^4 (1 - B/B_{c2})\times \cr\cr
 &&\times [1 +
1.21(1 - \sqrt{B/B_{c2}})^3]\lambda^{-2}[{\rm nm}^{-2}].
 \label{eq:Brandt_sigma_vs_H}
\end{eqnarray}

\begin{figure}[htb]
\includegraphics[width=1.03\linewidth]{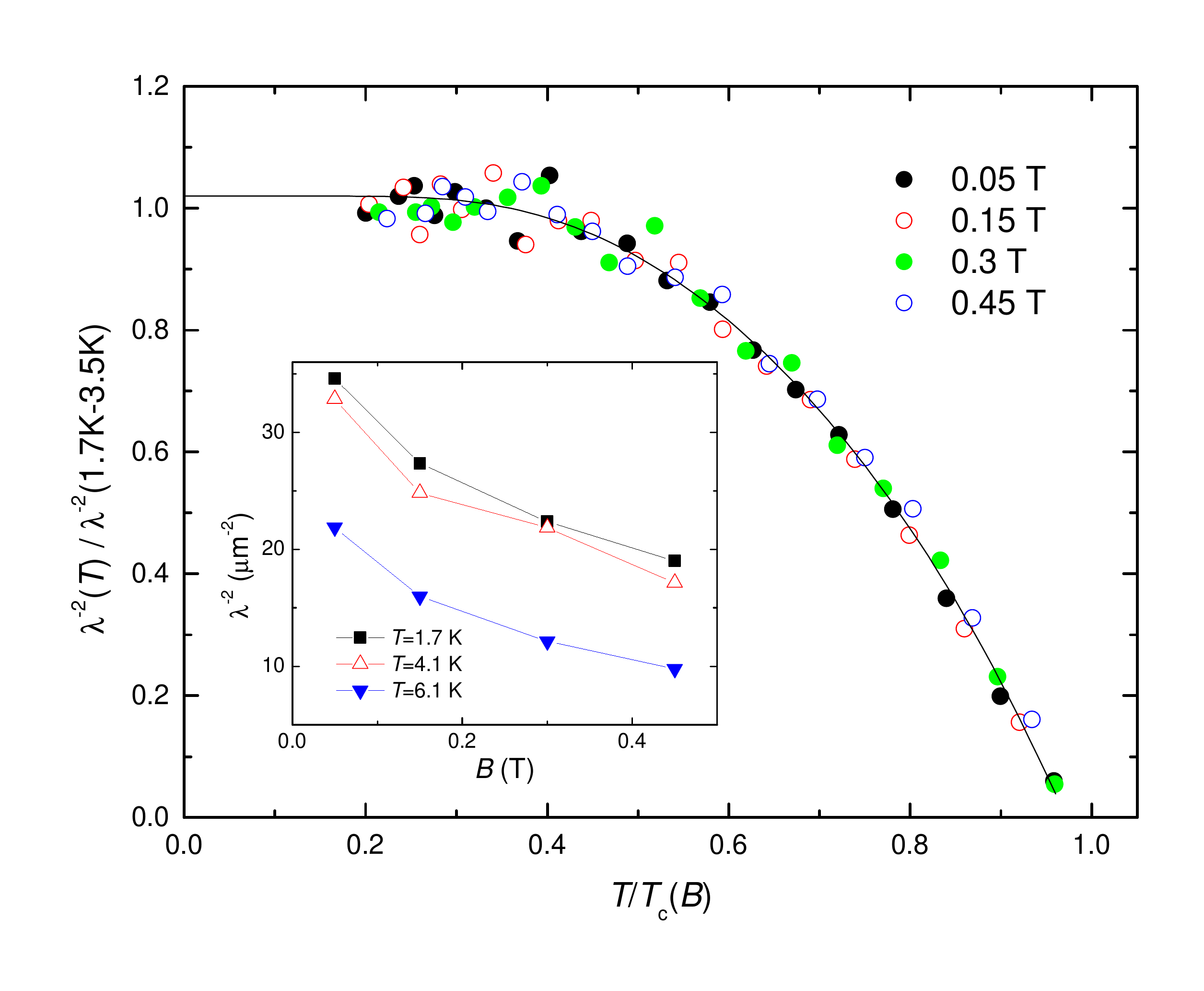}
\caption{(Color online) $\lambda^{-2}(T)$ normalized to its value averaged over the temperature range $1.7-3.5$~K  as a function of $T/T_c(B)$. The solid line is the fit by using the  weak-coupling BCS model (see Eq.~\ref{eq:lambda_0}). The inset shows the dependence of $\lambda^{-2}$ on the applied field  at $T=1.7$, 4.1 and 6.1~K.}
 \label{fig:lambda_eff}
\end{figure}

The insert of Fig.~\ref{fig:sigma_vs_T} shows the evolution of $\sigma_{sc}$ at $T=$1.7~K as a function of the applied magnetic field $B$. Each data point was obtained after cooling the sample in the corresponding field from above $T_c$ to 1.7~K. Under the assumption of field independent $\lambda$ the  dependence of $\sigma_{sc}$ on $B$ was analyzed by means of Eq.~(\ref{eq:Brandt_sigma_vs_H}) using the values of the upper critical field $B_{c2}$ as obtained in magnetization experiments [$B_{c2}(1.7)$~K$\simeq4.5$~T, see Fig.~\ref{fig:Hc2}]. It is clear from the inset of Fig.~\ref{fig:lambda_eff} that the theoretical $\sigma(B)$ is not in agreement with the data. If $B_{c2}$ is kept as a free parameter in the analysis, the fit yields $B_{c2}=1.6(1)$~T which is clearly inconsistent with the magnetization data. Therefore one has to conclude that the field independence of $\lambda$, which was implicitly assumed in Eq.(\ref{eq:Brandt_sigma_vs_H}), is not valid (the discussion on field dependence of $\lambda$ comes later in the paper).
The low-temperature value of $\lambda$ at $B=0$ [$\lambda(0,B=0)$] could be estimated by extrapolating two theory lines shown in the inset of Fig.~\ref{fig:sigma_vs_T} to $B=0$. This results in $\lambda(0,B=0)=155\pm10$~nm.

The temperature dependences of $\lambda^{-2}$ for $\mu_0H=0.05$, 0.15, 0.3, and 0.45~T was obtained from measured $\sigma_{sc}(T)$'s and $B_{c2}(T)$ by using Eq.~(\ref{eq:Brandt_sigma_vs_H}). Figure \ref{fig:lambda_eff} shows $\lambda^{-2}(T)$ normalized to its value averaged over the temperature range $1.7-3.5$~K as a function of $T/T_c(B)$. All data curves merge into the single line. The inset of Fig.~\ref{fig:lambda_eff} shows the field dependence of $\lambda^{-2}$ for $T=1.7$, 4.1 and 6.1~K.

As a first step we are going to discuss the temperature dependence of $\lambda^{-2}$. It is seen that below approximately one half of $T_c$,  $\lambda^{-2}$ is temperature independent. The solid line in Fig.~\ref{fig:lambda_eff} represents fit with the weak-coupling BCS model \cite{Tinkham75}:
\begin{equation}
\frac{\lambda^{-2}(T)}{\lambda^{-2}(0)}=
\frac{\rho_s(T)}{\rho_s(0)}= 1
+2\int_{\Delta(T)}^{\infty}\left(\frac{\partial f}{\partial
E}\right)\frac{E\ dE}{\sqrt{E^2-\Delta(T)^2}}.
 \label{eq:lambda_0}
\end{equation}
Here $\lambda^{-2}(0)$ and $\rho_s(0)$ are the  zero-temperature values of the magnetic penetration depth and the superfluid density, respectively, and $f=[1+\exp(E/k_BT)]^{-1}$ is the Fermi function. The temperature dependence of the gap is approximated by $\Delta(T)/\Delta(0)=\tanh\{1.82[1.018(T_c/T-1)]^{0.51}\}$ \cite{Khasanov05}, where $\Delta(0)$ is the maximum gap value at $T=0$. The fit results in $\Delta(0,B)/k_BT_c(B)=4.35(4)$, $\lambda^{-2}(T)/\lambda^{-2}(1.7-3.5$~K)=1.021(6), and $T/T_c(B)=0.972(3)$. For $T_c(B=0)\simeq8.4$~K (see Fig.~\ref{fig:Hc2}) we get $\Delta(T=0,B=0)=1.58(2)$~meV. Note that this value of the superconducting gap is close to $\Delta=1.85$~meV obtained from zero-field specific heat data by Takayama {\it et al.} \cite{Takayama12}.

It is noteworthy  that there is no need to introduce more that one gap parameter or to consider more complicated gap symmetry in order to satisfactorily describe $\lambda^{-2}(T)$ data. A fit using two superfluid density components with $s-$wave gaps $\Delta_1$ and $\Delta_2$: $ \lambda^{-2}(T)=\lambda_1^{-2}(T,\Delta_1)+\lambda_2^{-2}(T,\Delta_2) $,  as well a fit using an anisotropic $s-$wave gap function result in higher $\chi^2$ than obtained for the simple one gap s-wave model described above. From the analysis of $\lambda^{-2}(T)$ data alone one could therefore conclude that  SrPt$_3$P is a  a single band s-wave superconductor. Note that the similar conclusion was reached by Takayama {\it et al.} \cite{Takayama12} based on specific heat data. In the following we will suggest that this was a premature conclusion obtained without considering the field dependence of $\lambda$.

As follows from the inset in Fig.~\ref{fig:lambda_eff}, the field increase from 0.05 up to 0.45~T leads to decrease of $\lambda^{-2}$ by almost a factor of 2.
In a single band $s-$wave superconductors $\lambda$ is {\it independent} on the magnetic field \cite{Khasanov06,Kadono04,Khasanov05,Khasanov08}. A dependence of
$\lambda$ on $B$ is expected for superconductors containing nodes in the energy gap or/and multi-gap superconductors \cite{Khasanov07,Kadono04,Angst04,Bussmann-Holder07,Weyeneth10}. In the later case the superfluid density within one series of bands is expected to be suppressed faster by magnetic field than within the others \cite{Bussmann-Holder07,Weyeneth10}.

The single $s-$wave gap behavior of $\lambda^{-2}(T)$ (see Fig.~\ref{fig:lambda_eff} and the discussion above) and the multi-band features following after the upper critical field $B_{c2}$ and $\lambda^{-2}(B)$ measurements (Fig.~\ref{fig:Hc2} and the inset on Fig.~\ref{fig:lambda_eff}) allow us to assume that SrPt$_3$P is a {\it two-band} superconductor with energy gaps being {\it equal} within both bands.

Within a two-gap model the deviation from the simple field independence of $\lambda$ as well as the appearance of upward curvature of the upper critical field could  reflect the occurrence of two distinct coherence lengthes $\xi_1$ and $\xi_2$ for two bands (associated to the corresponding upper critical field values $B_{c2,i}=\phi_0/2\pi \xi^2_i$) \cite{Bussmann-Holder07,Weyeneth10, Gurevich_03, Cubitt03, Serventi04, Mansor_05}. For BCS superconductors the zero-temperature coherence length obeys the relation $\xi\propto \langle v_{F}\rangle/\Delta$, ($\langle v_F\rangle$ is the averaged value of the Fermi velocity). One could assume, therefore that in SrPt$_3$P the difference between $\xi_1$ and $\xi_2$ could be caused by the different Fermi velocities ($\langle v_{F,1}\rangle \neq \langle v_{F,2}\rangle$), while gaps remain the same ($\Delta_1=\Delta_2$). 

The statement about different $\langle v_{F}\rangle$'s in two Fermi surface sheets of SrPt$_3$P is fully confirmed by the calculated band structure  \cite{Nekrasov12, Kang12,Chen12, Subedi12}. According to Refs.~\onlinecite{Nekrasov12, Kang12,Chen12, Subedi12} there are two bands crossing the Fermi level having significantly different $ v_{F}$'s. The ratio of $v_F$'s is, {\it e.g.},  $\simeq2$ along $\Gamma-X$ and $\sim3-4$ along $\Gamma-Z$ directions of the  Brillouin zone.
It is worth to note that different Fermi velocities on the different superconducting bands suppose to be a common feature of multi-band superconductors as {\it e.g.} MgB$_2$ \cite{Choi02,Belashchenko01,Suderow04}, borocarbides \cite{Shulga_98_BC,Suderow04} , Fe-based supercondutors \cite{Evtushinsky09,Tamai10} {\it etc}.

\begin{figure}[htb]
\includegraphics[width=1.0\linewidth]{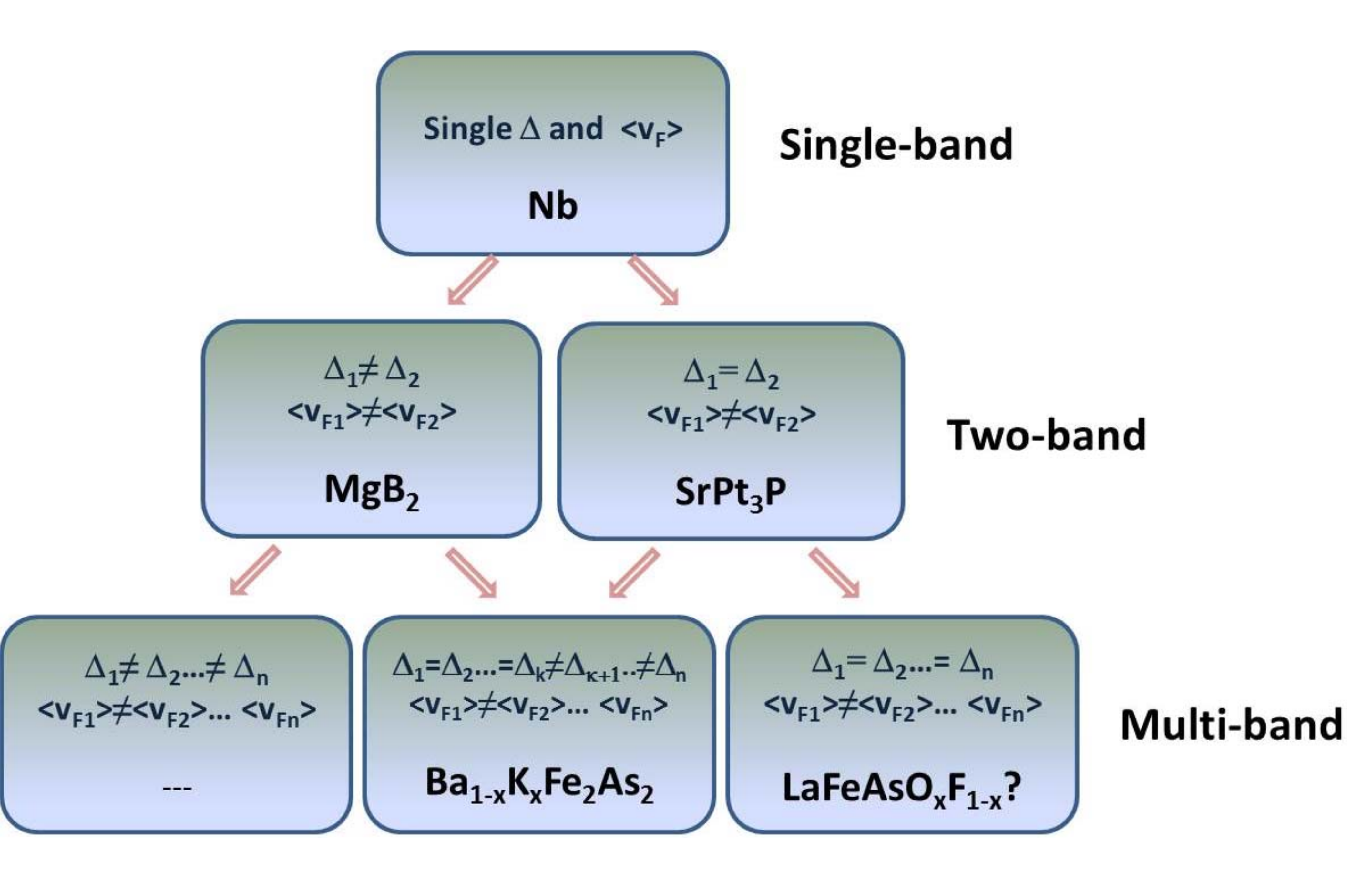}
\caption{(Color online) Schematic diagram representing relations between the various types of a single-band, two-band and multi-band superconductors. }
 \label{fig:Diagramm}
\end{figure}

Note SrPt$_3$P studied here is different from the most famous two-band superconductor MgB$_2$. In SrPt$_3$P the charge carriers in both bands suppose to be almost equally coupled to the phonons. Indeed, according to the band stricture calculations of Nekrasov {\it et al.} \cite{Nekrasov12} the carriers in two bands correspond to the relatively {\it similar} $p d \pi$ antibonding states of Pt(I)-P and Pt(II)-P ions, and are coupled to the {\it same} low-lying phonon modes confined on the ab plane. In MgB$_2$ only the $\sigma$ band carriers are coupled strongly to the so called E$_{2g}$ phonons, while the coupling of both, the $\sigma$ and the $\pi$, bands to the harmonic B$_{1g}$, A$_{2u}$, and E$_{1u}$ phonons is negligible \cite{Yildirim01}. We may conclude, therefore, that MgB$_2$ and SrPt$_3$P correspond to two limiting cases of two-band superconductivity with the energy gaps being nonequal ($\Delta_1\neq \Delta_2$, as in MgB$_2$) and equal ($\Delta_1=\Delta_2$, as in SrPt$_3$P). At the same time SrPt$_3$P remains the "true" two-band superconductor since, due to nonequal Fermi velocities ($\langle v_{F,1}\rangle \neq \langle v_{F,2}\rangle$), the carriers in various bands "respond" differently to the magnetic field (as shown here based on $B_{c2}(T)$ and $\lambda(B)$ studies and by  Takayama {\it et al.} \cite{Takayama12} based on the observation of nonlinear temperature behavior of the Hall resistivity).

 By following the above presented arguments we propose a schematic diagram describing relations between the single-, two-, and the multi-band superconductivity (see Fig.~\ref{fig:Diagramm}). The single-band superconductor has one gap and one averaged over the Fermi surface Fermi velocity ($\langle v_{F}\rangle$).  There are two type of two-band superconductors with energy gaps being equal ($\Delta_1= \Delta_2$) and nonequal ($\Delta_1\neq \Delta_2$). Both of these types are characterized, however, by nonequal $\langle v_{F}\rangle$'s. The "transition" from the two- to the multi-band superconductivity may occur by three different routes. (i) All gaps in all bands crossing the Fermi level are equal ($\Delta_1=\Delta_2 \ldots =\Delta_n$). This is probably the case for the optimally doped LaFeAsO$_{0.9}$F$_{0.1}$ having five Fermi surfaces (as most other Fe-based superconductors, see {\it e.g.} Ref.~\onlinecite{Johnston10} and references therein). As shown by Luetkens {\it et al.}~\cite{Luetkens08} the temperature evolution of the superfluid density of LaFeAsO$_{0.9}$F$_{0.1}$ is well described within the single $s-$wave gap approach, while  $\lambda^{-2}$ depends strongly on the magnetic field.
 It should be noted, however that the presence of two distinct gaps in LaFeAsO$_{0.9}$F$_{0.1}$ were reported by Gonnelli {\it et al.} \cite{Gonnelli09} based on the result of point contact Andreev reflection experiment. (ii) Gaps in some Fermi sheets are equal but in others are not ($\Delta_1=\Delta_2 \ldots =\Delta_k\neq\Delta_{k+1}\ldots\neq\Delta_n$).  A good example is the optimally doped Ba$_{1-x}$K$_x$Fe$_2$As$_2$ where three gaps are equal ($\simeq 9$~meV) while the last gap was found to be of approximately eight times smaller ($\simeq 1.1$~meV) \cite{Evtushinsky09,Khasanov09}. (iii) Gaps in all the Fermi sheets are different ($\Delta_1\neq\Delta_2\ldots\neq\Delta_n$).

To summaries, the temperature and the magnetic field dependence of the magnetic penetration depth $\lambda$ in SrPt$_3$P superconductor ($T_c\simeq 8.4$~K) were studied by means of muon-spin rotation. Below $T\simeq T_c/2$ the superfluid density $\rho_s\propto\lambda^{-2}$ is temperature independent which is consistent with a fully gapped superconducting state. The full $\rho_s(T)$ is well described within the single $s$-wave gap scenario with the zero-temperature gap value $\Delta=1.58(2)$~meV. At the same time $\lambda$ was found to be strongly field dependent which is the characteristic feature of the nodal gap and/or multi-band systems. The multi-band nature of the superconduicting state in SrPt$_3$P was further confirmed by observation of an upward curvature of the upper critical field. To conclude, all above presented results show  SrPt$_3$P to be a two-band superconductor with the equal gaps but different coherence  lengthes $\xi_i$ associated with the two Fermi surface sheets.

This work was performed at the Swiss  Muon Source (S$\mu$S), Paul Scherrer Institute (PSI, Switzerland).

\end{document}